\documentclass[tikz, border=10pt]{article}

\usepackage{ijcai26}

\usepackage{todonotes}

\usepackage{times}
\usepackage{soul}
\usepackage{url}
\usepackage[hidelinks]{hyperref}
\usepackage[utf8]{inputenc}
\usepackage[small]{caption}
\usepackage{graphicx}
\usepackage{amsmath}
\usepackage{amsthm}
\usepackage{booktabs}
\usepackage{algorithm}
\usepackage{algorithmic}
\usepackage[switch]{lineno}
\usepackage{tikz}
\usepackage{placeins}
\usetikzlibrary{shapes.geometric, arrows.meta, positioning, fit, calc, backgrounds, shadows}
\title{Democratizing Ski Safety: Real-Time Turn Segmentation with Smartphone IMU and Causal LSTM Networks}

\author{
Michał Szymocha\and
Piotr Kacprzak\and
Jakub Robak\And
Wojciech Turek \\
\affiliations
AGH University of Krakow, 30-059 Krakow, Poland\\
\emails
jakubrobak@agh.edu.pl
}

\begin{document}

\maketitle

\begin{abstract}

Anterior cruciate ligament (ACL) injury is one of the most common and serious injuries in sports, particularly among recreational skiers. Research shows that structured technique awareness and continuous feedback can significantly reduce the risk of such injuries, yet access to professional instructors is limited to wealthy athletes who can afford continuous private coaching, creating a harmful inequity in injury prevention. This gap can be mitigated by automating the real-time analysis of skiing techniques available to the wider recreational skiing community. The approach relies exclusively on inertial sensors embedded in standard smartphones, eliminating the need for specialized equipment and enabling broad social scalability. To support immediate feedback, the system operates causally, producing predictions based solely on past observations. The work is conducted in cooperation with professional ski instructors, ensuring that problem formulation, data annotation, and result evaluation reflect real-world coaching practices and injury prevention needs. The model is evaluated using Leave-One-Subject-Out validation on a public, in-the-wild dataset, demonstrating robust generalization across skiers, achieving an average directional accuracy of 89.8\%, while maintaining extremely low inference latency suitable for on-device mobile deployment. This work outlines a practical pathway to democratizing injury prevention in recreational sports.

\end{abstract}

%%%%%%%%%%%%%%%%%%%%%%%%%%%%%%%%%%%%%%%%%%%%%%%%%%%%%%%%%%%%%%%%%%
\section{Introduction}
\label{sec:intro}

Alpine skiing is one of the most popular forms of winter recreation. The latest industry reports estimate the number of slope visits at more than 366 million per year \cite{vanat2025international}. Its popularity remains unimpeded by a high risk of serious injury. Extensive studies of competitive skiing \cite{pujol2007incidence,jordan2017anterior} have shown a high incidence of anterior cruciate ligament (ACL) ruptures, highlighting the inherent biomechanical risks of this sport. In recreational skiing, these risks are further exacerbated by a lack of continuous technical supervision and awareness of movements, often leading to an unsafe loss of ski control \cite{ruedl2023association}. Beyond severe injuries, this constitutes a broad public safety challenge that affects millions of recreational skiers each season \cite{davey2019alpine}. Crucially, fundamental research has shown that awareness of structured technique and ongoing correction can significantly reduce the risk of serious knee sprains \cite{ettlinger1995method}.

Unlike professional athletes who train in structures with regular coaching support, recreational skiers typically lack access to systematic feedback on technique. Research on sport participation reveals persistent socioeconomic inequalities. Individuals of higher socioeconomic status are significantly more likely to participate in organized physical activity than those from lower income groups, indicating that financial resources influence access to sport \cite{richard2023socioeconomic}. Furthermore, research indicates that the costs associated with participating in sports are perceived as a significant barrier. Many families limit participation in activities due to their inability to purchase equipment, memberships, or training sessions \cite{o2025sport}. In addition, literature reviews highlight that equipment expenses, fees, and other costs are among the most common practical barriers to participation in sports across various age groups \cite{somerset2018barriers}. This creates a significant gap in safety support and movement education among the broader recreational ski community. Widely available smartphone-based mobile systems offering automated real-time monitoring could help alleviate this inequality by providing feedback without the recurring costs associated with traditional coaching.

Existing smartphone apps for skiers are usually based on GNSS data, which do not offer the time resolution required for detailed technique evaluation \cite{petrella2025accuracy}. 
There are advanced commercial systems, such as smart insoles with pressure sensors (e.g., Carv systems \cite{enoiu2023new}), that provide granular feedback. However, these are expensive and require specialized equipment, limiting their accessibility. 
In contrast, inertial sensors embedded in standard smartphones offer a widely available sensing modality for movement analysis, enabling socially scalable technique assessment \cite{azadi2022alpine}.

In order to become an efficient "digital safety assistant", a smartphone must first detect and understand the basic unit of a skier's movement, i.e. a single turn. Precise temporal segmentation, as shown in the literature \cite{martinez2021comprehensive}, is a necessary condition for further biomechanical analysis, which is crucial for safety assessment. Synchronization errors at this stage lead to false risk assessments, making a robust segmentation algorithm the foundation of the entire system.

Efficient operation in uncontrolled real-world conditions (in-the-wild) encounters a significant technological barrier resulting from the need to reconcile algorithmic robustness with real-time requirements. Physics-based methods that use rigid thresholds achieve high precision in laboratory settings but fail to generalize to uncontrolled conditions \cite{yamagiwa2014development,martinez2021comprehensive}. In contrast, advanced deep learning architectures often rely on non-causal or bidirectional processing, requiring access to future context \cite{chen2022energy} and thus precluding their use for real-time injury prevention. Therefore, there is an urgent need to develop an architecture that retains the robustness of deep learning while operating causally, eliminating the delays that result from analyzing future context.

This study is conducted in collaboration with professional ski instructors, ensuring that problem formulation, data annotation, and evaluation criteria reflect real-world coaching, movement education, and safety practices rather than purely algorithmic assumptions. As a result, this work provides a foundational building block for future smartphone-based digital safety assistants by enabling a reliable real-time interpretation of ski maneuvers.

The contributions of this work are threefold.

(i) We formulate the problem of ski turn segmentation as a causal frame-level recognition task suitable for real-time processing on mobile devices.

(ii) We propose a lightweight, learning-based approach for turn segmentation using only inertial sensors embedded in standard smartphones, designed to operate robustly in real-world, in-the-wild conditions.

(iii) We validate the proposed approach under subject-independent Leave-One-Subject-Out evaluation on a public, expert-annotated dataset, demonstrating strong generalization across skiers while meeting the latency constraints required for on-device deployment.

%%%%%%%%%%%%%%%%%%%%%%%%%%%%%%%%%%%%%%%%%%%%%%%%%%%%%%%%%%%%%%%%%%

\section{Related Work}

Traditional turn detection methods in alpine skiing are based on physical models. A comprehensive review of published approaches \cite{martinez2021comprehensive} shows that existing methods achieve high accuracy in the laboratory, but field validation revealed reduced performance for non-carving styles (e.g., drifted: 0.833, snowplow: 0.538 for the Ratio score) and variability
dependent on snow conditions \cite{martinez2019development}. So far, machine learning has focused mainly on classifying skiers' entire runs, not on the precise segmentation of turn boundaries\cite{jones2016automatic}.

A recent study \cite{robak2025turn} addressed the problem of detecting ski turns using smartphone IMU sensors by developing a gradient descent-based algorithm on a public dataset. The method achieved an F1 score of 0.943 in detecting complete turns (turn-level) and 89.7\% accuracy in mapping the turn direction (left/right) for each signal time point. However, this approach requires offline processing of entire runs (post-hoc optimization), operates at the level of global transition points instead of frame-level prediction, and is evaluated using randomized data splitting without subject hold-out validation.

In related domains, LSTM networks have shown promising results. LSTM models for gait phase estimation have achieved errors below 100 ms on IMU sensors \cite{ding2018real,tang2024imu}, and recent studies using a single IMU on a smartphone reported a median error of around 70-80 ms with 92\% accuracy
of events \cite{larsen2024accurate}. Human activity recognition (HAR) using LSTM and CNN achieves 95--99\% accuracy \cite{xia2020lstm}, and video temporal action segmentation uses bidirectional models achieving high precision in boundary localization \cite{ding2023temporal}.

Despite these advances, existing approaches fundamentally differ from the requirements of ski turn segmentation in real-world recreational conditions. Most HAR and temporal segmentation systems operate in time windows of a few seconds rather than single-sample predictions, utilize bidirectional architectures requiring access to future data, and are predominantly validated in controlled laboratory settings. These assumptions limit the applicability of such methods to real-time, safety-oriented systems running on mobile devices in uncontrolled skiing conditions. 

Furthermore, compact LSTM models can perform real-time causal inference on mobile devices \cite{chen2022energy}, but their application to temporal event detection on raw IMU data from ski turns in real-world terrain (in-the-wild) remains unexplored.

Based on this review, four key research gaps can be identified:

(i) The absence of causal methods for real-time ski turn segmentation using mobile inertial sensors.

(ii) Limited generalization between skiers, skill levels, and riding styles in real uncontrolled slope conditions, where existing algorithms exhibit reduced performance.

(iii) The lack of efficient deployment of advanced sequence models, such as LSTM networks, on resource-constrained smartphones for real-time temporal event detection.

(iv) The absence of systematic comparisons between simple rule-based approaches and modern deep learning architectures for the ski turn segmentation task.

This work addresses these gaps by proposing a causal LSTM architecture for real-time ski turn segmentation on raw IMU data from standard smartphones in real-world, in-the-wild conditions, based on an extended dataset with additional attributes (skill level, skiing style) to improve cross-skier generalization.
%%%%%%%%%%%%%%%%%%%%%%%%%%%%%%%%%%%%%%%%%%%%%%%%%%%%%%%%%%%%%%%%%%
\section{Dataset and Expert Domain Knowledge}
\label{sec:Dataset}
The effectiveness of AI systems for recreational safety support depends on the reliability of training data and its connection with the biomechanics of movement. This section describes a set of IMU data recorded in real-world conditions (in-the-wild) that was analyzed and annotated by an expert ski instructor. By combining smartphone-based sensing with instructor-provided annotations and contextual metadata, the dataset supports safety-oriented applications such as risk-aware feedback and technique assessment in recreational sports. Importantly, relying on commodity mobile devices reduces economic barriers and enables inclusive access to these systems, regardless of financial means or access to professional instruction.

\subsection{Data Source}

The experimental basis of this work is a publicly available IMU dataset recorded during real-world skiing runs (in-the-wild)  \cite{robak2025turn}. The dataset is annotated by a certified ski instructor using synchronized video recordings that serve as a reference for temporal alignment and the definition of the turn boundary. In accordance with the annotation protocol, each timestamp is assigned a binary label corresponding to the current turn direction (left or right), as described in professional curricula, such as the PSIA Alpine Technical Manual. In this setting, straight or no-turn segments are treated as brief transition periods rather than separate target states, since our goal is to analyze the active turning phases most relevant for turn-shape assessment and safety feedback. The collection comprises 105 runs performed by 11 different skiers: 7 males and 4 females, with 9 participants aged 18 to 30 years and 2 aged 40 to 55 years, resulting in 1,781 recorded turns. Each ski run is recorded as a continuous multivariate IMU time series sampled at 10 Hz using inertial sensors embedded in standard smartphones. Following prior work on smartphone-based ski motion analysis, the device is mounted on the skier’s calf, providing stable coupling with lower-limb kinematics while remaining practical for real-world use. The participants represented a full spectrum of skill levels, from beginner to certified ski instructor, making the data highly representative of the diverse recreational skier population. An important element of the realism of the data is the fact that the skiers moved freely, without an imposed route or gates, which introduces natural variability in the trajectory and movement dynamics typical of ski tourism. The data was recorded using inertial sensors built into standard smartphones, which guaranties high social scalability of the solution.

\subsection{Expert-Led Metadata Enrichment}

A key value of the collection is the rigorous process of determining the "ground truth" conducted by a ski instructor. Instead of relying on simplistic physical thresholds, the limits of each turn were defined on the basis of biomechanical analysis of synchronized video recordings. This high labeling precision is essential for ACL injury prevention systems, where fractions of a second in motion phase detection can strongly affect the accuracy of biomechanical risk assessment in dynamically changing slope conditions \cite{sporri2016sidecut}.

To further adapt the data to the requirements of real-world injury prevention, the original dataset was augmented with expert metadata. The Enrichment process was conducted by the same certified ski instructor, who performed a detailed analysis of synchronized video recordings and raw inertial sensor signals. As a result, each of the 105 runs was assigned precise contextual attributes: skier skill level (skier level) distributed as 2 beginners, 5 intermediates, 3 advanced, and 1 expert; skiing style (skier style); and a unique user identifier (skier ID). In accordance with the annotation protocol, each timestamp is assigned a binary label corresponding to the current turn direction (left or right), without introducing an explicit straight-skiing or no-turn class.

The addition of these parameters allows for rigorous validation of the model's ability to generalize across different user technical profiles. From the perspective of ACL injury prevention, this approach is essential for creating a system that can accurately interpret movement dynamics in individuals with a wide range of technical skills, from beginners to experts. Integrating domain knowledge in the feature set definition stage lays the foundation for the democratization of professional safety systems, making them effective for every smartphone user.

%%%%%%%%%%%%%%%%%%%%%%%%%%%%%%%%%%%%%%%%%%%%%%%%%%%%%%%%%%%%%%%%%%

\section{Methodology}
\label{sec:method}

This section presents a causal approach to ski turn recognition from smartphone IMU data. The goal is accurate frame-level segmentation in real-world conditions under mobile and real-time constraints. We use a lightweight recurrent model that relies only on past observations.

\subsection{Problem Formulation}
We address the problem of online ski turn recognition from multivariate inertial sensor data. Given a continuous IMU time series recorded during a skiing run, the task is to predict the current turn direction at each time step, resulting in a frame-level binary labeling of left and right turns.

Formally, let $x_t \in {\rm I\!R} ^d$ denote the multichannel IMU measurement at time step $t$. 
The objective is to learn a model $f_\theta$ that produces an online prediction of the current turn direction based solely on past and present observations in \autoref{eq:prediction}.
\begin{equation}
\hat{y}_t = f_\theta(x_{1:t}).
\label{eq:prediction}
\end{equation}

The problem is challenging due to substantial noise in consumer-grade IMU signals, pronounced inter-subject variability in skiing technique, and changing environmental conditions encountered in real-world skiing. Moreover, enforcing causality introduces an inherent trade-off between temporal precision and latency, as reliable detection of turn transitions may require the accumulation of sufficient historical context without ever accessing future data, which is a strict constraint for both the unidirectional LSTM and preprocessing imposed by real-time mobile deployment rather than accuracy optimization.

\subsection{Data Preprocessing}

The preprocessing pipeline is designed to stabilize raw IMU signals for recurrent modeling while preserving causality and minimizing additional latency, which is critical for real-time operation on mobile devices.

To handle phase discontinuities in rotational signals, we apply phase unwrapping to the orientation components to obtain continuous angular representations suitable for temporal modeling \cite{robak2025turn}. Without unwrapping, angular representations exhibit artificial discontinuities that introduce spurious high-magnitude jumps, which can significantly degrade temporal modeling in recurrent networks.

To reduce high-frequency measurement noise while preserving short-term motion dynamics, each signal component is filtered online using an exponential moving average (EMA). This filter is strictly causal, as each output value depends solely on the current observation and the previous filter state, allowing it to be applied in real time without access to future samples. EMA filtering is performed independently for each run to prevent information leakage between sessions.

Then, feature normalization is subsequently applied using statistics computed exclusively on the training data within each fold.

To prepare data for temporal modeling, we generate fixed-length sequences using a sliding window approach with a unit stride, labeled according to the ground-truth class of the final timestamp, which is consistent with the causal setting adopted for online prediction. This solution provides sufficient time context to model skiing dynamics while limiting the sequence length for stable optimization and effective training.

In addition to raw sensor signals, we derive a small set of physics-informed features to capture motion characteristics relevant to skiing behavior. For accelerometer and gyroscope signals, we compute magnitude vectors using the $\ell_2$ norm to represent overall motion intensity. Jerk is computed as the temporal derivative of acceleration, providing a measure of movement smoothness \cite{hogan2009sensitivity}. These features offer significant physical context with negligible computational overhead. By capturing motion intensity and smoothness through the $\ell_2$ norm and jerk, we provide the model with interpretable indicators of skier stability and dynamics that are not immediately accessible from raw signals alone. Most importantly, we calculate and include the first and second derivatives of yaw. Finally, local rolling statistics are computed to capture short-term temporal patterns without violating causality.

Overall, the proposed data preprocessing pipeline provides noise-robust, strictly causal input representations suitable for real-time recursive modeling in uncontrolled, real-world skiing conditions.

\subsection{Data Augmentation}

The data augmentation strategy is introduced to mitigate limited subject diversity and variability in execution speed, which are particularly pronounced under subject-independent (LOSO) evaluation and can hinder generalization in real-world deployment. Notably, the dataset is relatively small, and models tend to overfit quickly on it.

To increase temporal variability while preserving the semantic structure of individual turns, we apply data augmentation using segment-wise time warping. Unlike standard window warping applied to entire sequences, we perform warping independently within each annotated turn segment of a ski run. This design preserves behavioral transitions while allowing realistic variations within turns.

Window warping is applied using four temporal scale factors: 0.5, 0.75, 1.25, and 1.5, following established practice in inertial time-series augmentation \cite{window-warping}. The window ratio parameter is set to 1.0, allowing warping to span the full duration of each turn segment. This procedure produces four augmented variants for each original run.

The warping operation relies on temporal interpolation, which has the additional side effect of mildly smoothing high-frequency sensor noise while preserving low-frequency motion dynamics. As a result, the augmented signals lead to more stable optimization during training. 

All augmented samples are generated exclusively from the training subset within each LOSO fold, and models are trained using the augmented data. It improves the generalization ability of a model, thanks to the mild smoothing side effect of augmentation. Importantly, the original (non-augmented) training data is \emph{not} used during optimization. Validation and test subsets remain completely unmodified, ensuring that hyperparameter tuning and final evaluation reflect performance on realistic, non-distorted sensor dynamics.

In general, this augmentation strategy effectively quadruples the size of the training set while maintaining physically plausible temporal relationships within behavioral segments. It also reflects realistic differences in how quickly a skier executes a turn without altering the underlying motion pattern.

\subsection{Training Strategy}

A critical failure mode in kinematics-based Human Activity Recognition (HAR) is the "identity bias", where models overfit biometric characteristics (e.g., unique sensor mounting angles or specific posture habits) rather than learning universal movement mechanics. Standard randomized $k$-fold cross-validation exacerbates this by allowing temporally correlated windows from the same subject to appear in both training and testing sets, leading to inflated performance estimates. \cite{loso}

To enforce the learning of generalizable subject-invariant physical features, we employ a strict Leave-One-Subject-Out (LOSO) training strategy. The training process is iterative: for each unique subject with at least 5 descents $k \in \{1, \dots, N\}$, we create a test set containing exclusively the runs performed by subject $k$. The model is trained from scratch on the remaining pool $S \setminus \{k\}$. This strategy simulates a "cold-start" deployment scenario, ensuring that the optimization process never observes the kinematic characteristics of the target user, thereby testing the model's ability to apply learned biomechanics to unseen individuals.

Additionally, within each training iteration, we enforce strict data hygiene to prevent spatiotemporal leakage. The training pool is further divided into train and validation subsets. Crucially, this split is performed at the entire run rather than at the window level. Splitting by window would result in training and validation samples that are milliseconds apart, trivially correlating in phase and amplitude. Our file-level splitting ensures that the validation set represents distinct events.

Finally, to stabilize the training of the recurrent architecture, we adopt a Dense Supervision objective. Rather than optimizing a single prediction at the end of the sequence, which forces error gradients to backpropagate throughout the temporal depth, often leading to vanishing gradients, we supervise the network at every time step $t$. This approach enhances our model's causal capabilities. The objective function is Cross-Entropy Loss.

Synthesizing strict LOSO validation with dense supervision and run-level hygiene yields a robust optimization pipeline that effectively eliminates data leakage while stabilizing the learning of subject-invariant biomechanics.

%%%%%%%%%%%%%%%%%%%%%%%%%%%%%%%%%%%%%%%%%%%%%%%%%%%%%%%%%%%%%%%%%%
% \newpage
\section{Proposed Architecture}
\begin{figure}[H]
\centering
\includegraphics[width=0.41\linewidth]{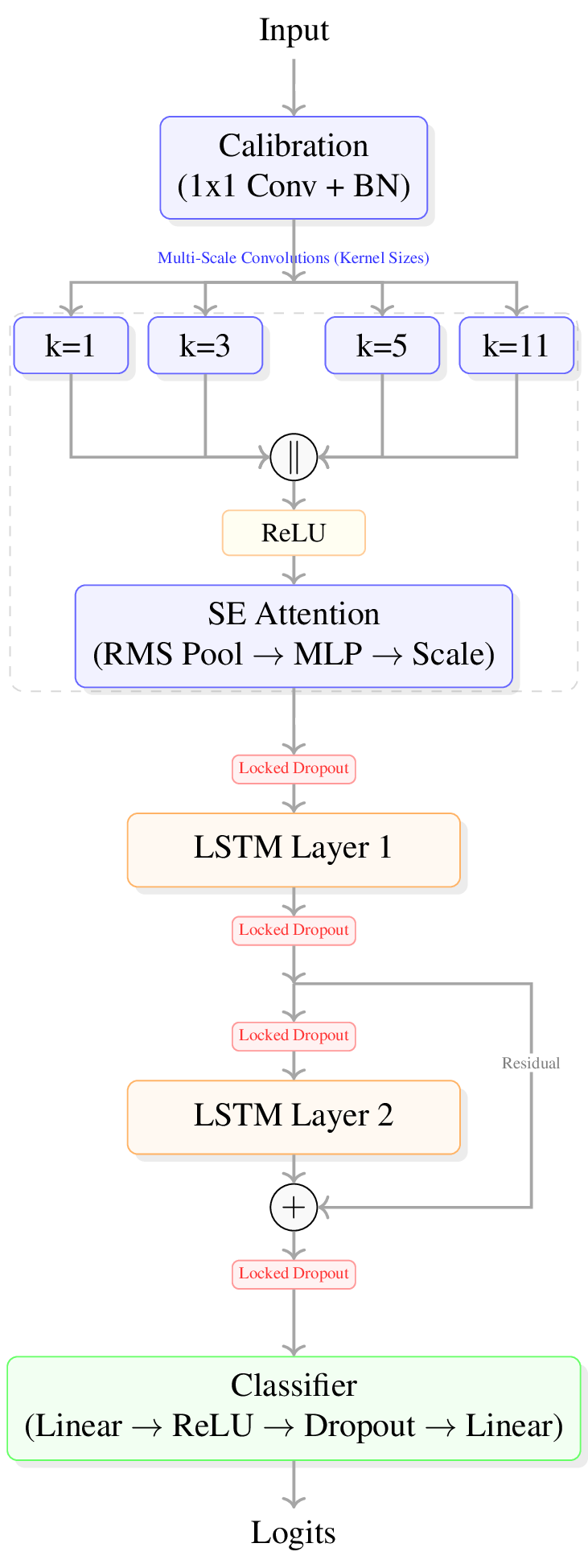}

\caption{SkiC-LSTM architecture scheme}
\label{figure:SkiC-LSTM}
\end{figure}

To address the challenges of alpine skiing recognition, specifically the variability in sensor placement, the multi-frequency nature of motion, and long-term temporal dependencies—we propose a hybrid Residual Recurrent Network. Our architecture transforms raw inertial signals into a rich latent representation through three distinct stages, as illustrated in \autoref{figure:SkiC-LSTM}.

    \paragraph{Learnable Sensor Calibration and Alignment} 
    Sensor misalignment remains a bottleneck in wearable-based motion analysis, where physical orientation often varies between subjects. Rather than relying on imprecise manual alignment, we incorporate an entry-level Calibration Layer designed to normalize the input manifold before temporal processing. This $1 \times 1$ convolutional block serves as a learnable spatial transformer, optimizing a projection matrix $W_{cal}$ that aligns disparate sensor frames into a canonical feature representation. This data-driven realignment reduces the model’s sensitivity to physical setup, allowing it to focus on the underlying motion dynamics rather than subject-specific sensor bias.

    \paragraph{Multi-Scale Spatiotemporal Feature Extraction} 
    Skiing dynamics exhibit features at different temporal resolutions: high-frequency vibrations indicating ski-snow interaction (edge chatter) and low-frequency distinct phases (weight shift, steering). A single convolutional kernel size fails to capture this spectrum simultaneously. We employ a Multi-Scale Convolutional Block comprising four parallel branches with kernel sizes $k \in \{1, 3, 5, 11\}$. 
    \begin{itemize}
        \item The $k=1$ branch captures instantaneous global trends.
        \item The $k=3, 5$ branches extract short-term interactions and rapid orientation changes.
        \item The $k=11$ branch captures long-duration maneuvers and trajectory shapes.
    \end{itemize}
    The outputs are concatenated to form a unified feature map. To explicitly model the interdependencies between physical channels (e.g., the correlation between roll rate and lateral acceleration), we integrate a Squeeze-and-Excitation (SE) block \cite{squeeze-excitation}. This mechanism recalibrates channel-wise feature responses by explicitly modeling dependencies, effectively acting as a soft attention mechanism that suppresses noise and amplifies informative physical signals.

    \paragraph{Residual Recurrent Encoding} 
    We model the dynamics of each turn using a stacked LSTM architecture. To improve training stability, we incorporate residual skip connections ($h_t^l = \text{LSTM}(h_t^{l-1}) + h_t^{l-1}$), which allow gradients to flow more freely and enable the model to bypass layers that do not contribute to the representation. To ensure these representations remain robust, we use Locked Dropout. Unlike standard dropout, this approach fixes the dropout mask across the entire sequence, preventing the model from relying on localized noise and encouraging the learning of stable temporal patterns \cite{merity2017regularizing}.

\begin{table}[H]
\centering
\resizebox{\linewidth}{!}{
\begin{tabular}{lcl}
\toprule
\textbf{Layer} & \textbf{Output Dim.} & \textbf{Configuration / Parameters} \\ 
\midrule
\multicolumn{3}{l}{\textit{1. Pre-processing}} \\
Permute & $(C_{in}, T)$ & Transpose $(B, T, C_{in}) \to (B, C_{in}, T)$ \\
Calibration & $(C_{in}, T)$ & Conv1d ($C_{in} \to C_{in}$, $k=1$), bias=False \\
BatchNorm & $(C_{in}, T)$ & BN1d ($C_{in}$ features), $\epsilon=1e^{-5}$ \\ 
\midrule
\multicolumn{3}{l}{\textit{2. Feature Extraction}} \\
Multi-Scale Conv & $(64, T)$ & 4 Parallel Branches ($C_{out}=16$ each): \\
& & \hspace{3mm} 1. Conv1d ($k=3, p=1$) \\
& & \hspace{3mm} 2. Conv1d ($k=5, p=2$) \\
& & \hspace{3mm} 3. Conv1d ($k=11, p=5$) \\
& & \hspace{3mm} 4. Conv1d ($k=1, p=0$) \\
SE Block & $(64, T)$ & Reduction $r=4$, RMS Pooling \\
& & MLP: $64 \to 16 \xrightarrow{\text{ReLU}} 64 \xrightarrow{\text{Sigmoid}}$ \\
Activation & $(64, T)$ & ReLU \\ 
\midrule
\multicolumn{3}{l}{\textit{3. Recurrent Layers}} \\
Permute & $(T, 64)$ & Transpose to $(B, T, C_{feat})$ \\
Dropout & $(T, 64)$ & Locked Dropout ($p=0.30$) \\
LSTM Layer 1 & $(T, 128)$ & Input: 64, Hidden: 128, Unidirectional \\
LSTM Layer 2 & $(T, 128)$ & Input: 128, Hidden: 128, Unidirectional \\
& & \textit{Residual Connection}: $x + \text{LSTM}(x)$ \\ 
\midrule
\multicolumn{3}{l}{\textit{4. Classification Head}} \\
Slice & $(128)$ & Select last time step $t=T$ \\
FC Layer 1 & $(128)$ & Linear $128 \to 128$, ReLU \\
Dropout & $(128)$ & Standard Dropout ($p=0.30$) \\
FC Layer 2 & $(2)$ & Linear $128 \to 2$ (Logits) \\ 
\bottomrule
\end{tabular}
}
\caption{Detailed architecture of the proposed SkiC-LSTM model. The input sequence has dimensions $B \times T \times C_{in}$, where $T$ represents the window size and $C_{in}$ represents the number of input features. Model parameters were tuned using grid search over a candidate set of hyperparameters.}
\label{tab:model_arch}
\end{table}
\FloatBarrier

\section{Experiments}
\label{sec:setup}

Multiple experiments were conducted to examine the performance of our architecture in comparison to existing approaches under standard evaluation protocols and the generalization across different skiers, skier levels, and styles. Comparisons are performed under identical training and testing conditions against strong baselines.

\subsection{Compared Baselines}
We compare the proposed method against a set of classical and deep-learning baselines commonly used in time-series Human Activity Recognition (HAR), covering both feature-based and end-to-end learning paradigms. 

Concretely, we include \textbf{Random Forest} and \textbf{XGBoost} as representative feature-based baselines, as tree ensembles are routinely reported in HAR benchmarks and comparative studies \cite{Syed2020_LogisticsHAR_RF_XGBoost,ZhangZhaoLi2019_XGBoostIndoorHAR}.

For end-to-end learning, we use \textbf{LSTM} and \textbf{ResNet1D}-style architectures, which are commonly adopted deep baselines for wearable/smartphone time-series HAR across standard datasets \cite{Hammerla2016IJCAI_HAR,OrdonezRoggen2016_DeepConvLSTM,KimKim2025_1DResNet_HAR}. While bidirectional models are feasible, we do not include them because they require future context and therefore fall outside the real-time causal setting considered here.

\subsection{Evaluation Metrics}
\label{sec:eval-metrics}
Performance is evaluated using Accuracy and Intersection over Union. Each time step is treated as an independent prediction, measuring the model’s ability to maintain consistent semantic interpretation throughout the duration of a turn, rather than only predicting the dominant label of a window. Accuracy quantifies the proportion of timestamps for which the model correctly predicts the direction of a turn. The Intersection over Union metric is calculated by taking the average of the best possible \textit{IoU} for each turn matched with the most overlapping continuous prediction.

We compute accuracy metrics both globally and per individual skier. This dual-level evaluation allows us to verify the overall efficacy of the dataset while simultaneously ensuring the model remains robust against inter-subject variability in style and technique. To avoid the high uncertainty associated with skiers who have very few timestamps, we reduced the evaluation pool to valid skiers with at least 5 descents. This resulted in a final test dataset consisting of 99 runs, 9 different skiers, and 1656 turns.

The inclusion of IoU is particularly important for this domain. While accuracy provides a general performance overview, it can be misleadingly high in datasets dominated by a background class (e.g., long periods of straight skiing). IoU, conversely, measures the precise temporal overlap between the predicted and ground-truth events. It penalizes both over-prediction (false positives) and under-prediction (missed duration) equally, ensuring that a high score reflects not just the detection of a turn but the accurate recovery of its boundaries.

This evaluation protocol is deliberately strict: even small temporal misalignments at turn boundaries are penalized at every affected timestamp. Consequently, the reported metrics reflect both classification correctness and temporal precision, which are critical for downstream applications such as real-time feedback and fine-grained skill assessment.

\subsection{Implementation Details}
All models operate on sliding windows of fixed length $W=12$ with a stride of one. This dense overlapping scheme ensures high temporal resolution and maximizes data utilization while preserving local temporal continuity.

All available IMU channels are used as input features; however, some of them are transformed, allowing the models to exploit the full inertial signal space captured by the device.

Neural network models are implemented in \texttt{PyTorch}~\cite{pytorch}. The ResNet1D baseline follows a standard residual design consisting of an initial convolutional stem (kernel size 3), followed by six residual blocks operating at a constant channel width of 128. Each block uses kernel size 3 and dropout rate of 0.1.

Random Forest and XGBoost baselines are trained using standard fitting procedures from scikit-learn, while their hyperparameters were tuned via grid search across commonly used configuration ranges. All baselines share the same windowing strategy and are evaluated under identical data partitions to ensure a fair comparison.

SkiC-LSTM follows the specification in \autoref{tab:model_arch}. We optimize it with AdamW using learning rate $1 \cdot 10^{-5}$ and weight decay $1 \cdot 10^{-4}$. To improve generalization, we apply a dropout rate of $0.3$. Training runs for a maximum of 40 epochs, utilizing an early stopping mechanism with a patience of 5 epochs.

A characteristic of frame-level classifiers is that the raw output exhibits high-frequency jitter. While these short duration anomalies do not significantly impact global accuracy metrics, they affect the visual smoothness required for user-facing applications. To mitigate this, we implemented a median filter as a post-processing step. Although this introduces a 0.5s latency, resulting in a near-real-time rather than a strictly real-time system, it effectively eliminates artifacts. Crucially, this lightweight post-processing, combined with the SkiC-LSTM’s efficient inference, ensures the pipeline remains computationally inexpensive, making it highly suitable for direct deployment on mobile edge devices where resource constraints are paramount. For comparison purposes, we applied the same post-processing to every baseline to ensure that all of the results are comparable.

All preprocessing and data-dependent transformations are computed per training fold, as described in Training Strategy section, with no test-set statistics used. All baselines are evaluated under the same causal input protocol as our model (tree models use an unrolled representation). Experiments use the dataset from \autoref{sec:Dataset}.

\section{Results}
\label{sec:results}

The proposed method is compared with representative baselines under identical training and testing splits, and performance is reported using evaluation metrics defined in \autoref{sec:eval-metrics}. The quantitative results are presented first, followed by ablation studies and analyzes to better understand the contributions of individual components.

\subsection{Quantitative Results}
\begin{table}[H]
    \centering
    \resizebox{\linewidth}{!}{
    \begin{tabular}{lcccc}
        \toprule
        Method & Per-skier Acc & Per-run Acc &  Per-run IoU \\
        \midrule
        ResNet1D & 86.45 & 87.18 & 75.36 \\
        Standard LSTM & 86.84 & 87.30 & 75.64 \\
        Random Forest & 87.18 & 87.73 & 75.01 \\
        XGBoost & 88.50 & 88.66 & 78.51 \\
        \textbf{SkiC-LSTM (Ours)} & \textbf{89.77}  & \textbf{89.85} & \textbf{79.45} \\
        \bottomrule
    \end{tabular}}
    \caption{Segmentation performance (\%) on the Test Set.}
    \label{tab:results}
\end{table}
\FloatBarrier

SkiC-LSTM achieves the best results under identical splits, with the highest per-skier accuracy (89.77), per-run accuracy (89.85), and per-run IoU (79.45) in \autoref{tab:results}. The IoU gain indicates more temporally coherent segments and reduced boundary jitter.

\begin{table}[h]
    \centering
    \small 
    \resizebox{\linewidth}{!}{%
    \begin{tabular}{lcccc}
        \toprule
        Method \textbackslash \ Style & Carving & Quick & Skidded & Snowplow \\
        \midrule
        ResNet1D         & 87.83 & 78.89 & 87.89 & 85.38 \\
        LSTM             & 89.08 & 84.44 & 87.97 & 82.57 \\
        RandomForest     & 89.40 & 82.52 & 87.80 & 85.46 \\
        XGBoost          & 90.94 & \textbf{88.18} & 89.30 & 84.74 \\
        \textbf{SkiC-LSTM (Ours)} & \textbf{92.10} & 87.60 & \textbf{89.70} & \textbf{88.24} \\
        \bottomrule
    \end{tabular}%
    }
    \caption{Comparison of Accuracy (\%) across different skiing styles.}
    \label{tab:style_acc_comparison}
\end{table}

\begin{table}[h]
    \centering
    % \small
    \resizebox{\linewidth}{!}{
    \begin{tabular}{lcccc}
        \toprule
        Method \textbackslash \ Level & 0 & 1 & 2 & 3 \\
        \midrule
        ResNet1D              & 83.58 & 88.91 & 89.33 & 80.58 \\
        LSTM                  & 82.00 & 89.68 & 90.32 & 81.02 \\
        RandomForest          & 85.12 & 88.97 & 89.10 & 84.01 \\
        XGBoost               & 84.45 & 90.45 & 90.82 & 86.31 \\
        \textbf{SkiC-LSTM (Ours)} & \textbf{87.07} & \textbf{90.99} & \textbf{91.53} & \textbf{88.11} \\
        \bottomrule
    \end{tabular}}
    \caption{Comparison of Accuracy (\%) across different skier skill levels (0: Beginner, 1: Intermediate, 2: Advanced, 3: Expert).}
    \label{tab:weighted_acc_levels}
\end{table}
\FloatBarrier
\begin{table}[h]
    \centering
    \small 
    \label{tab:skier_acc_models}
    \resizebox{\linewidth}{!}{
    \begin{tabular}{lccccc}
        \toprule
        SkierID & ResNet1D & LSTM & RF & XGB & SkiC-LSTM \\
        \midrule
        1  & 89.13 & \textbf{91.54} & 91.11 & 90.89 & 91.44 \\
        2  & 82.00 & 78.25 & 83.67 & 81.63 & \textbf{85.58} \\
        4  & 91.36 & 92.06 & 91.32 & 91.71 & \textbf{92.31} \\
        5  & 85.09 & 85.60 & 86.50 & 87.14 & \textbf{88.49} \\
        6  & 88.31 & 87.95 & 87.17 & \textbf{89.79} & 89.55 \\
        8  & 83.35 & 85.18 & 82.55 & 88.20 & \textbf{89.22} \\
        9  & 89.04 & 90.11 & 88.69 & 89.19 & \textbf{92.15} \\
        10 & 89.18 & 89.90 & 89.63 & \textbf{91.68} & 91.06 \\
        11 & 80.58 & 81.02 & 84.01 & 86.31 & \textbf{88.11} \\
        \midrule
        \textbf{Mean} & 86.45 & 86.84 & 87.18 & 88.50 & \textbf{89.77} \\
        \textbf{$\mathbf{\sigma}$} & 3.78 & 4.78 & 3.26 & 3.21 & \textbf{2.20} \\
        \bottomrule
    \end{tabular}}
    \caption{Comparison of Accuracy (\%) for each skier with at least 5 runs.}
\end{table}
\FloatBarrier

We further stratify the results by skiing style, skill level, and individual skiers to assess generalization and subject-specific variability.

\subsection{Robustness Analysis}

Visual inspection confirms the architecture's strong capacity for preserving the morphological shape of motion sequences. Quantitative error analysis reveals that accuracy losses are primarily attributable to temporal phase shifts rather than event misclassifications; the model remains highly reliable in detecting turn occurrence and duration. These phase shifts typically manifest as consistent differences of $n$ frames, a phenomenon inherent to the causal nature of SkiC-LSTM processing historical context.

Regarding style generalization, the model demonstrated the expected superior predictive performance on clean carving turns. For "snowplow" techniques, data were scarce: only 10 descents, mostly from one subject during LOSO training. Despite this, the performance degradation was acceptable, with a directional accuracy drop of only 2\% \autoref{tab:style_acc_comparison}. In particular, our model demonstrated better generalization compared to recurrent baselines, which failed to converge effectively in low-data scenarios.

Performance stratification by skier level followed an expected trend: advanced skiers yielded the highest accuracy, followed by the intermediate group, which accounted for the majority of the dataset. Consistent with our expectations, \autoref{tab:weighted_acc_levels} shows that the model's performance was lowest for beginners.  This drop is attributable to a combination of factors: the inherently high variability of beginner movements and the limited sample size, as only two beginner skiers were included in the study.

Finally, the IoU gains in \autoref{tab:results} indicate improved segment overlap with the ground truth, reflecting better boundary alignment and reduced temporal fragmentation.

Our model achieved an inference time of around 0.1 ms per frame (on a single core 3 GHz CPU) leaving several orders of magnitude of compute headroom for deployment on resource-constrained devices, while matching the performance of much heavier offline models \cite{robak2025turn}, which are not suitable for real-time feedback.

%%%%%%%%%%%%%%%%%%%%%%%%%%%%%%%%%%%%%%%%%%%%%%%%%%%%%%%%%%%%%%%%%%

\section{Conclusion}
\label{sec:conclusion}

The developed hybrid LSTM architecture successfully segments IMU data into individual skiing maneuvers with high efficiency. By achieving precise turn segmentation even in the presence of "in-the-wild" noise, the model proves to be more resilient than traditional physics-based approaches, which often struggle with the irregularities of real-world recreational skiing.

The causal architecture and very low inference time support real-time processing on standard smartphone hardware. By enabling automated data structuring without the need for professional oversight, this work offers a practical path to mitigate harmful inequity in access to coaching and injury prevention tools. Ultimately, this research provides the essential framework for a future real-time smartphone assistant, where reliable maneuver segmentation serves as the critical first step toward comprehensive technique evaluation and skier safety\footnote{Source code, pretrained weights, and curated dataset are available at \url{https://github.com/mszymocha/SkiC-LSTM}.}.

\newpage
\bibliographystyle{named}
\bibliography{ijcai26}

\end{document}